\documentclass[11pt]{article}
\usepackage{amsmath}
\usepackage{amsfonts}
\usepackage{amssymb}
\usepackage{graphicx}
\usepackage{bm}
\usepackage{color}
\usepackage{amscd}

\def\bea{\begin{eqnarray}}
\def\eea{\end{eqnarray}}

\begin{document}
\begin{center}
\LARGE {\bf On the new version of Generalized Zwei-Dreibein Gravity }
\end{center}

\begin{center}
{M. R. Setare \footnote{E-mail: rezakord@ipm.ir}\hspace{1mm} ,
H. Adami \footnote{E-mail: hamed.adami@yahoo.com}\hspace{1.5mm} \\
{\small {\em  Department of Science, University of Kurdistan, Sanandaj, Iran.}}}\\

\end{center}

\begin{center}
{\bf{Abstract}}\\
In this paper we consider a generalization of zwei-dreibein gravity with a chern-Simons term associated to a constraint term which fixed the torsion. We count the local degrees of freedom of this model using Hamiltonian analysis and show that in contrast to the usual GZDG which has 2 bulk local degrees of freedom, our model has 3 propagating modes. Then by looking at the quadratic Lagrangian, we determine that these propagating modes are 3 massive graviton with different masses. Finally we obtain AdS wave solution as an example solution for this model.

\end{center}

\section{Introduction}
Pure Einstein-–Hilbert gravity in three dimensions exhibits no propagating physical degrees of freedom \cite{2',3'}. But adding the gravitational Chern-Simons term produces a
propagating massive graviton \cite{4'}. A few years ego \cite{6} a new theory of massive gravity (NMG) in three dimensions
has been proposed. This theory is
 equivalent to the three-dimensional
Fierz-Pauli action for a massive spin-2 field at the linearized level, moreover NMG  is parity
invariant. As a result, the gravitons acquire the same mass for both
helicity states, indicating two massive propagating degrees of freedom. Usually the theories including the
terms given by the square of the curvatures have the massive spin 2
mode and the massive scalar mode in addition to the massless
graviton. Also the theory has ghosts due to negative energy
excitations of the massive tensor. The unitarity of NMG
 was discussed in \cite{d1,d2} (see also \cite{dan11,dan12, oh1}). Although, it has been shown the
compliance of the NMG with the holographic c-theorem \cite{9,10}, NMG has a bulk-boundary
unitarity conflict. In another term either the bulk or the boundary theory is non-unitary, so there is a clash between the positivity of the two Brown-Henneaux boundary $c$ charges
 and the bulk energies \cite{11}. There is this possibility to extend NMG to higher curvature theories. One of these extension
 of NMG has been done by Sinha \cite{9} where he has added the $R^3$ terms to the action. The other modification is the extension to the Born-Infeld type action \cite{12}. But these extensions of NMG  did not solve
the unitary conflict \cite{9,12,13}. The recently constructed Zwei Dreibein Gravity (ZDG) shows that there is a viable
alternative to NMG \cite{1,2}. ZDG contain two Dreibeine, and is a two-derivative model. The authors of \cite{4} have obtained the Chern-Simons-like formulation of NMG from ZDG model by field and parameter redefinitions. When one linearize ZDG about $AdS_3$ background it propagates two massive helicity-2 modes. ZDG model with $\beta_2=0$ is free from Boulwar-Deser ghost, but in the case $\beta_2\neq0$, this model has ghost \cite{27}. If one demand that a linear combination of the dreibeine to be invertible, then ZDG will be free of ghost. A parity violating generalisation of ZDG (which we call GZDG) can be obtained by a combination of $L_{ZDG}(\beta_2=0)$ plus Lorentz-Chern-Simons (LCS)term. From \cite{2} we know that GZDG in contrast with ZDG is free of Boulwar-Deser ghost at all. In the present paper we add a constraint term to the Lagrangian of GZDG for fixing torsion and introduce ${GZDG}^{+}$ model.\\ Our paper is organized as follows. In section 2 we introduce our model and obtain the field equations. Then in section 3, we study the Hamiltonian analysis of the ${GZDG}^{+}$ model and obtain the number of local degrees of freedom. We show that there are 3
propagating modes. To see whether they are ghosts or
tachyons, we need look at the quadratic action. So in section 4, we do linearized analysis. We show one can avoided from ghost with imposing some conditions on the parameters of the model. In section 5 we study the AdS waves solutions propagating on AdS$_{3}$ background. We conclude in section 6 with a discussion of the our results.

\section{Generalization of Zwei-Dreibein Gravity}
The Lagrangian 3-form of Zwei-Dreibein gravity is given by \cite{1}
\begin{equation}\label{1}
  \begin{split}
     L_{ZDG}=-M_{P}  & \{ \sigma e_{1} \cdot R_{1} + e_{2} \cdot R_{2}+ \frac{m^{2}}{6} \left( \alpha _{1} e_{1} \cdot e_{1} \times e_{1} +\alpha _{2} e_{2} \cdot e_{2} \times e_{2} \right) \\
       & -\frac{1}{2} m^{2} \left( \beta _{1} e_{1} \cdot e_{1} \times e_{2} +\beta _{2} e_{2} \cdot e_{1} \times e_{2} \right) \},
  \end{split}
\end{equation}
where two Lorentz vector valued one-forms $ e_{I}^{\hspace{1.5 mm} a} $ ($I=1,2$) are dreibein \footnote{In this paper we use from the notation \cite{3}.}. Also $\omega _{I} ^{\hspace{1.5 mm} a}$ are dualised spin-conections and dualized curvature two-form define in terms of $\omega _{I}$ as $ R_{I}=d \omega _{I}+\frac{1}{2}\omega_{I} \times \omega_{I} $. As we know, the torsion two-form define as $T_{I}=D_{I}e_{I}=de_{I}+\omega_{I} \times e_{I}$, where $D_{I}$ is the exterior covariant derivative with respect to $\omega_{I}$. It should note that the dualized curvature two-form and the torsion two-form satisfy Bianchi identities:
\begin{equation}\label{2}
D_{I}R_{I}=0, \hspace{1.5 cm} D_{I}T_{I}=R_{I} \times e_{I}.
\end{equation}
In paper \cite{2}, the authors have proposed a  Lagrangian 3-form for generalized zwei-dreibein gravity (GZDG),
 \begin{equation}\label{3'}
L_{GZDG}=L_{ZDG}(\beta _{2}=0)+\frac{M_{P}}{2 \mu} ( \omega_{1} \cdot \omega_{1} +\frac{1}{3} \omega_{1} \cdot \omega_{1} \times \omega_{1} )
\end{equation}
Here, we add a constraint term to the above Lagrangian for fixing torsion. Thus, we write the Lagrangian as follow
\begin{equation}\label{3}
L_{GZDG}^{+}=L_{ZDG}(\beta _{2}=0)+\frac{M_{P}}{2 \mu} ( \omega_{1} \cdot \omega_{1} +\frac{1}{3} \omega_{1} \cdot \omega_{1} \times \omega_{1} ) + M_{P} h \cdot T_{1},
\end{equation}
where $h$ is an auxiliary Lorentz vector valued one-form which play the role of Lagrange multiplier in this action. One can read off the field equations from the above Lagrangian as
\begin{equation}\label{4}
\sigma R_{1}+\frac{m^{2}}{2} \alpha _{1} e_{1} \times e_{1} - m^{2} \beta _{1} e_{1} \times e_{2}-D_{1}h=0,
\end{equation}
\begin{equation}\label{5}
\sigma T_{1}-\frac{1}{\mu}R_{1} -  e_{1} \times h=0,
\end{equation}
\begin{equation}\label{6}
R_{2}+\frac{m^{2}}{2} \alpha _{2} e_{2} \times e_{2} - \frac{m^{2}}{2} \beta _{1} e_{1} \times e_{1}=0,
\end{equation}
\begin{equation}\label{7}
T_{1}=0,
\end{equation}
\begin{equation}\label{8}
T_{2}=0.
\end{equation}
Now, we take $g_{1 \mu \nu}$ as the physical metric and assume that $e_{1}^{\hspace{1.5 mm} a}$ is invertible then one can expressed $e_{2}^{\hspace{1.5 mm} a}$ in terms of $e_{1}^{\hspace{1.5 mm} a}$ and its derivatives,
\begin{equation}\label{9}
e_{2 \hspace{1.5 mm} \mu}^{\hspace{1 mm} a}=\frac{\alpha _{1}}{2 \beta _{1}}e_{1 \hspace{1.5 mm} \mu}^{\hspace{1.5 mm} a} +\frac{\sigma}{m^{2} \beta_{1}} S_{1 \hspace{1.5 mm} \mu}^{\hspace{1.5 mm} a} +\frac{1}{\mu m^{2} \beta _{1}}C_{1 \hspace{1.5 mm} \mu}^{\hspace{1.5 mm} a},
\end{equation}
where
\begin{equation}\label{10}
 S_{1 \mu \nu}=\mathcal{R}_{1 \mu \nu}-\frac{1}{4} g_{1 \mu \nu} \mathcal{R}_{1} , \hspace{1 cm} C_{1 \mu \nu}= \epsilon _{1 \mu}^{\hspace{3 mm} \alpha \beta} \nabla _{1 \alpha} S_{1 \nu \beta},
\end{equation}
are the Schouton tensor and the Cotton tensor, respectively. In the above definition $\mathcal{R}_{1 \mu \nu}$, $\mathcal{R}_{1}$ and $\epsilon _{1 \mu}^{\hspace{3 mm} \alpha \beta}$ are Ricci curvature, Ricci scalar and levi-civita tensor, respectively. So that all of these tensors compute in terms of $g_{1 \mu \nu}$. Comparing \eqref{9} with corresponding result in the \cite{4} we see that the last term added to ZDG result with $\beta _{2}=0$. Using the equation \eqref{7} one can find $\omega _{1}=\omega_{1}(e_{1})$. Now, we write $\omega_{2}$ as a power series in $1/m^{2}$
\begin{equation}\label{11}
\omega_{2}=\sum_{n=0}^{\infty } \frac{1}{m^{2n}} \Omega^{(2n)}.
\end{equation}
By putting this expression into \eqref{8} we find that
\begin{equation}\label{12}
\Omega^{(0)a}=\omega _{1}^{\hspace{1.5 mm} a}, \hspace{1 cm} \Omega^{(2)a}=-\frac{2}{\alpha _{1}}(\sigma C_{1}^{\hspace{1.5 mm} a}+\frac{1}{\mu}  H_{1}^{\hspace{1.5 mm} a} ),
\end{equation}
and
\begin{equation}\label{13}
\Omega^{(2k)a}=-\frac{2}{\alpha _{1}} \varepsilon^{bcd} \varepsilon _{ijk} (\delta ^{a} _{d} e_{1}^{\hspace{1.5 mm} k}-\frac{1}{2} \delta^{k}_{d} e_{1}^{\hspace{1.5 mm} a} ) \Omega ^{(2k-2) \hspace{1.5 mm} i}_{\hspace{8 mm} b}(\sigma S_{1c}^{\hspace{2.5 mm} j}+\frac{1}{\mu} C_{1c}^{\hspace{2.5 mm} j} ),\hspace{1 mm} k \geq 2,
\end{equation}
where $ H_{1 \mu \nu}=\epsilon _{1 \mu}^{\hspace{3 mm} \alpha \beta} \nabla _{1 \alpha} C_{1 \nu \beta} $. Hence, one can calculate $R_{2}$ as a power series in $1/m^{2} $ and gets
\begin{equation}\label{14}
R_{2}^{(0)}=R_{1}, \hspace{1 cm} R_{2}^{(2)}=D_{1} \Omega ^{(2)}=-\frac{2}{\alpha _{1}} (\sigma D_{1} C_{1} + \frac{1}{\mu} D_{1} H_{1}),
\end{equation}
and
\begin{equation}\label{15}
R_{2}^{(2k)}=D_{1} \Omega ^{(2k)}+\frac{1}{2} \sum_{i=1}^{k-1} \Omega ^{(2i)} \times \Omega ^{(2k-i)} ,\hspace{3 mm} k \geq 2.
\end{equation}
Substituting these expressions into \eqref{6}, we obtain a differential equation for $ e_{1}^{\hspace{1.5 mm} a} $ as a power series in $ 1/m^{2} $.

\section{Hamiltonian analysis}
 Since GZDG model is a Chern-Simones like theory, so in this section we apply  the Hamiltonian analysis of Chern-Simones like theory which fully investigated in \cite{2} and then we find the number of degrees of freedom of GZDG model with a torsion constraint. As we know, the generic Lagrangian 3-form for a Chern-Simons like theory is given by
\begin{equation}\label{16}
L=\frac{1}{2} g_{rs} a^{r} \cdot da^{s}+\frac{1}{6} f_{rst} a^{r} \cdot a^{s} \times a^{t},
\end{equation}
where $ a^{ra}=a^{ra}_{\hspace{3 mm} \mu} dx^{\mu} $ are Lorentz vector valued one-forms, $r$ ($r=1,...,N$) and $a$ refer to flavour and Lorentz indices, respectively. Also, $g_{rs}$ is a symmetric constant metric on the flavour space and $f_{rst}$ is a totally symmetric "flavour tensor" which interprate as the coupling constans. One can separate space and time by rewriting $a^{ra}$ in the form $a^{ra}=a^{ra}_{\hspace{3 mm} 0} dx^{0}+a^{ra}_{\hspace{3 mm} i} dx^{i}$. Then we can write the Lagrangian 3-form as $L=\mathcal{L}dx^{0} \wedge dx^{1} \wedge dx^{2} $, where
\begin{equation}\label{17}
\mathcal{L}=-\frac{1}{2} g_{rs} \varepsilon ^{ij} a^{r}_{\hspace{1.5 mm} i} \cdot \dot{a}^{s}_{\hspace{1.5 mm} j}-a^{r}_{\hspace{1.5 mm} 0} \cdot \phi _{r} .
\end{equation}
In the above formula the dote on top of $a^{s}_{\hspace{1.5 mm} j}$ denote the time derivative, and $\varepsilon ^{ij}=\varepsilon ^{0ij}$, also $\phi _{r}^{\hspace{1.5 mm} a}$ define as
\begin{equation}\label{18}
\phi _{r}^{\hspace{1.5 mm} a}=-\varepsilon ^{ij} [ g_{rs} \partial _{i} a^{sa}_{\hspace{2 mm} j}+\frac{1}{2} f_{rst} (a^{s}_{\hspace{1.5 mm} i} \times a^{t}_{\hspace{1.5 mm} j})^{a} ].
\end{equation}
Thus, one can find the Hamiltonian as
\begin{equation}\label{19}
\mathcal{H}=a^{r}_{\hspace{1.5 mm} 0} \cdot \phi _{r}.
\end{equation}
Since the Lagrangian is independent of $\dot{a}^{r}_{\hspace{1.5 mm} 0}$,  we can interpret $\phi _{r}^{\hspace{1.5 mm} a}=0$ as the primary constraints and $a^{ra}_{\hspace{2 mm} 0}$ as the Lagrange multipliers. From the Lagrangian \eqref{16} one can find that the Poisson brackets of the canonical variables are
\begin{equation}\label{20}
\{ a^{r}_{\hspace{1.5 mm} ai}(x),a^{s}_{\hspace{1.5 mm} bj}(x^{\prime}) \}_{P.B.}=\varepsilon _{ij} g^{rs} \eta _{ab} \delta ^{(2)}(x-x^{\prime}).
\end{equation}
We can define the "smeared" functions $\varphi [\xi]= \int_{\Sigma}d^{2}x \xi^{r}_{\hspace{1.5 mm} a}\phi _{r}^{\hspace{1.5 mm} a}$, where $\xi^{r}_{\hspace{1.5 mm} a}$ is a test function and $\Sigma$ is constant $t$ space-like hypersurface. One can add a term to the smeared functions for defining new functions $\Phi[\xi]=\varphi[\xi]+Q[\xi]$ so that their variation with respect to $a^{r}_{\hspace{1.5 mm} ai}$ not provided boundary terms. In this manner, we can calculate Poisson brackets of the constraint functions which are given by
\begin{equation}\label{21}
\{ \Phi[\alpha],\Phi[\beta] \}_{P.B.}=\Phi[[\alpha,\beta]]-\int_{\Sigma} d^{2}x \alpha^{r}_{\hspace{1.5 mm} a} \beta ^{s}_{\hspace{1.5 mm} b} (\mathcal{P}^{ab})_{rs} - \int_{\partial \Sigma} dx^{i} g_{rs} \alpha^{r} \cdot \partial _{i} \beta ^{s},
\end{equation}
where
\begin{equation}\label{22}
(\mathcal{P}^{ab})_{rs}=f^{t}_{\hspace{1.5 mm} q[r } f_{s]pt} \eta ^{ab} \Delta ^{pq} + 2 f^{t}_{\hspace{1.5 mm} r[s} f_{q]pt} (V^{ab})^{pq} ,
\end{equation}
and
\begin{equation}\label{23}
\Delta ^{pq}=\varepsilon ^{ij} a^{p}_{\hspace{1.5 mm} i} \cdot a^{q}_{\hspace{1.5 mm} j}, \hspace{1.5 cm} (V^{ab})^{pq} = \varepsilon ^{ij} a^{pa}_{\hspace{2.5 mm} i} a^{qb}_{\hspace{2.5 mm} j} .
\end{equation}
One can choose $\xi ^{r}_{\hspace{1.5 mm} a}(x) $ in a manner so that the boundary integrals vanishes. Then using \eqref{21} and definition of $\Phi[\xi]$, we find the Poisson brackets of the primary constraints
\begin{equation}\label{24}
\{ \phi _{r}^{\hspace{1.5 mm} a }(x),\phi _{s}^{\hspace{1.5 mm} b }(x^{\prime}) \} _{P.B.}=\delta ^{(2)}(x-x^{\prime}) [ f^{t}_{\hspace{1.5 mm} rs } \varepsilon ^{ab} _{\hspace{2.5 mm} c} \phi _{t}^{\hspace{1.5 mm} c }(x) - (\mathcal{P}^{ab})_{rs}] .
\end{equation}
In \cite{2}, the authors have argued that the consistency conditions which guarantee time-independence of the primary constraints are equivalent to a set of  "integrability conditions" which can be derived from the equations of motion, and these give us secondary constraints. One can easily obtain the equations of motion from the Lagrangian \eqref{16},
\begin{equation}\label{25}
g_{rs} da^{sa}+\frac{1}{2}f_{rst} (a^{s} \times a^{t})^{a}=0 ,
\end{equation}
Taking exterior derivative from the equations of motion, we have
\begin{equation}\label{26}
f^{t}_{\hspace{1.5 mm} q[r } f_{s]pt} a^{ra} a^{p} \cdot a^{q}=0.
\end{equation}
If we demand that one of the our 1-form fields, for example $a^{ra}=k^{a} $, be invertible and the sum over $r$ is non-zero for only $k^{a}$, then by separating space and time part of this 2-form, we obtain secondary constraints as follow
\begin{equation}\label{27}
\psi _{I}=F_{I,pq} \Delta ^{pq}, \hspace{1.5 cm} I=1,...,M,
\end{equation}
where $F_{I,pq}=-F_{I,qp}$. Now we can calculate Poisson brackets of secondary and primary constraints, so we obtain following Poisson brackets
\begin{equation}\label{28}
\{ \Phi[\xi],\psi _{I} \}_{P.B.}=2 \varepsilon ^{ij} (F_{I,rp} \partial _{i} \xi ^{r} \cdot a^{p}_{\hspace{1.5 mm} j } +f^{t}_{\hspace{1.5 mm} rs } F_{I,pt} \xi ^{r} \cdot a^{s}_{\hspace{1.5 mm} i } \times a^{p}_{\hspace{1.5 mm} j } ),
\end{equation}
then we find that
\begin{equation}\label{29}
\{ \phi _{r}^{\hspace{1.5 mm} a }(x) ,\psi _{I}(x^{\prime}) \}_{P.B.}=2 \varepsilon ^{ij} \delta ^{(2)}(x-x^{\prime}) (-F_{I,rp} \partial _{i} a^{pa}_{\hspace{2.5 mm} j } +f^{t}_{\hspace{1.5 mm} rs } F_{I,pt} (a^{s}_{\hspace{1.5 mm} i } \times a^{p}_{\hspace{1.5 mm} j })^{a} ).
\end{equation}
Also, one can find Poisson brackets of secondary constraints as
\begin{equation}\label{30}
\{ \psi _{I}(x),\psi _{J}(x^{\prime}) \}_{P.B}=-4 F_{I,pq} F_{J,rs} g^{qs} \Delta ^{pr} \delta ^{(2)}(x-x^{\prime}).
\end{equation}
We mention here an important result of \cite{2} which state that one can count the number of local degrees of freedom by following formula
\begin{equation}\label{31}
D= 6 N - 2 \times (3 N - rank ( \mathcal{P} ) - M )-1 \times ( rank( \mathcal{P} ) + 2 M )=rank ( \mathcal{P} ).
\end{equation}
It should noted that this formula is valid when Poisson brackets of secondary constraints all vanishes.\\
Now, we will apply this procedure to determine the number of local degrees of freedom of our suggested model. In this model, there are five flavours of 1-form, $a^{ra}=(e_{1},e_{2},\omega _{1}, \omega _{2},h)$. By comparing the Lagrangian GZDG \eqref{3} and the Lagrangian \eqref{16} one can read off the flavour metric and flavour tensor,
\begin{equation}\label{32}
\begin{split}
    & g_{e _{1}\omega _{1}}=-\sigma , \hspace{1 cm} g_{e _{2}\omega _{2}}=-1 , \hspace{1 cm} g_{e _{1} h}=1 , \hspace{1 cm} g_{\omega _{1}\omega _{1}}=\frac{1}{\mu} \\
     & f_{e_{1} \omega _{1} \omega _{1}} = -\sigma , \hspace{6 mm} f_{e_{2} \omega _{2} \omega _{2}} = -1 , \hspace{6 mm} f_{e_{1} e _{1} e _{1}} = -m^{2} \alpha _{1} , \hspace{6 mm} f_{e_{2} e _{2} e _{2}} = -m^{2} \alpha _{2} \\
     & \hspace{1 cm} f_{e_{2} e _{1} e _{1}} = m^{2} \beta _{1} , \hspace{1 cm} f_{e_{1} \omega _{1} h} = 1 , \hspace{1 cm} f_{\omega _{1} \omega _{1} \omega _{1}} = \frac{1}{\mu}
\end{split}
\end{equation}
We can write down \eqref{26} for this model as follow
\begin{equation}\label{33}
\begin{split}
    & e _{1}^{\hspace{1.5 mm} a} e _{1} \cdot e _{2}=0, \hspace{1 cm} e _{1}^{\hspace{1.5 mm} a} e _{1} \cdot h=0, \hspace{1 cm} e _{1}^{\hspace{1.5 mm} a} e _{1} \cdot (\omega _{1}-\omega _{2})=0, \\
     & (\omega _{1} ^{\hspace{1.5 mm} a} -\omega _{2} ^{\hspace{1.5 mm} a})e _{1} \cdot e _{2}+\frac{\mu}{m^{2} \beta _{1} } e _{1} \cdot h + e_{2} ^{\hspace{1.5 mm} a} e _{1} \cdot (\omega _{1} -\omega _{2})=0.
\end{split}
\end{equation}
If we demand that $e _{1}^{\hspace{1.5 mm} a}$ be invertible then we will have three secondary constraints:
\begin{equation}\label{34}
\psi _{1}= \Delta ^{e_{1}e_{2}}, \hspace{1 cm} \psi _{2}= \Delta ^{e_{1} h}, \hspace{1 cm} \psi _{3}= \Delta ^{e_{1} \omega _{1}}-\Delta ^{e_{1} \omega _{2}}.
\end{equation}
Now we can read off $F_{I,pq}$ and then calculate \eqref{30} and eventually we have $\{ \psi _{I}(x),\psi _{J}(x^{\prime}) \}_{P.B}=0$. Therefore we are allowed to use \eqref{31} for counting the number of local degrees of freedom of this model. The only thing we must know is rank of $\mathcal{P}$. By using \eqref{32} and the secondary constraints \eqref{34}, we calculate the $\mathcal{P}$ matrix which expressed in the formula \eqref{22}:
\begin{equation}\label{35}
\frac{\mathcal{P}}{m^{2} \beta _{1}}= \left( \begin{smallmatrix}
     -2V_{[ab]}^{\omega _{1}e_{2}}+2V_{[ab]}^{e_{2}\omega _{2}}-\frac{\mu}{m^{2}\beta _{1}}V_{ab}^{hh} & V_{ab}^{e_{2}e_{1}} & \frac{\mu}{m^{2}\beta _{1}} V_{ab}^{h e_{1}} & V_{ab}^{\omega _{1}e_{1}}-V_{ab}^{\omega _{2}e_{1}} & -V_{ab}^{e_{2}e_{1}} \\
                V_{ab}^{e_{1}e_{2}} & 0 & 0 & -V_{ab}^{e_{1}e_{1}} & 0 \\
                \frac{\mu}{m^{2}\beta _{1}}V_{ab}^{e_{1}h} & 0 & -\frac{\mu}{m^{2}\beta _{1}}V_{ab}^{e_{1}e_{1}} & 0 & 0 \\
                V_{ab}^{e_{1}\omega _{1}}-V_{ab}^{e_{1}\omega _{2}} & -V_{ab}^{e_{1}e_{1}} & 0 & 0 & V_{ab}^{e_{1}e_{1}} \\
                -V_{ab}^{e_{1}e_{2}} & 0 & 0 & V_{ab}^{e_{1}e_{1}} & 0 \\
   \end{smallmatrix}\right),
\end{equation}
This is a $15 \times 15$ matrix, and the rank of this matrix is $6$ (one can use of maple or mathematica softwares for calculating its rank) and then the number of local degrees of freedom is $6$,
\begin{equation}\label{36}
D= 6 \times 5 - 2 \times 6-1 \times  12 =6.
\end{equation}
Notice that $e_{2 \mu \nu}$, $h_{\mu \nu}$ and ($\omega _{1 \mu \nu}-\omega _{2 \mu \nu}$), by virtue of \eqref{33} and invertibility of $e_{1}$ , are symmetric. Now the question is: are there ghosts or not? So we should determine the type of degrees of freedom. We will answer to this question in the next section.

\section{Linearized analysis}
We suppose that $\bar{e}^{a}$ and $\bar{\omega}^{a}$ are dreibein and dualized spin-connection for AdS$_3$ background which has  negative cosmological constant $\Lambda = - \frac{1}{l^{2}}$. Then we can take $e_{1}^{\hspace{1.5 mm} a}=\bar{e}^{a}$, $e_{2}^{\hspace{1.5 mm} a}=\gamma \bar{e}^{a}$, where $\gamma$ is just a scaling parameter, and $\omega _{I}^{\hspace{1.5 mm} a}=\bar{\omega}^{a}$ as the background solution of the model, and these are solves equations of motion \eqref{4}-\eqref{8} provided that the following equations satisfied
\begin{equation}\label{37}
\alpha _{1}=2 \gamma \beta _{1}+\frac{\sigma}{m^{2} l^{2}} , \hspace{1.5 cm} \alpha _{2} \gamma ^{2}=\beta _{1}+\frac{1}{m^{2} l^{2}}.
\end{equation}
We now expand two dreibeins, two dualized spin-connections and the auxiliary 1-form field $h^{a}$ about  AdS$_3$ background as follow
\begin{equation}\label{38}
\begin{split}
    & e_{1}^{\hspace{1.5 mm} a}=\bar{e}^{a}+ \kappa u_{1}^{\hspace{1.5 mm} a}, \hspace{1 cm} e_{2}^{\hspace{1.5 mm} a}=\gamma (\bar{e}^{a}+ \kappa u_{2}^{\hspace{1.5 mm} a}),\hspace{1 cm} \omega _{1}^{\hspace{1.5 mm} a}=\bar{\omega}^{a}+\kappa v_{1}^{\hspace{1.5 mm} a}, \\
     & \omega _{2}^{\hspace{1.5 mm} a}=\bar{\omega}^{a}+\kappa v_{2}^{\hspace{1.5 mm} a}, \hspace{1 cm} h^{a}=\frac{1}{2 \mu l^{2}} (\bar{e}^{a}+ \kappa u_{1}^{\hspace{1.5 mm} a})+\kappa p^{a} ,
\end{split}
\end{equation}
where $\kappa$ is a small expansion parameter. By substituting these expressions into the Lagrangian \eqref{3}, and using \eqref{37} which cancels the linear term in the expansion of the Lagrangian, we will find that the quadratic Lagrangian for the fluctuations $u_{I}^{\hspace{1.5 mm} a}$, $v_{I}^{\hspace{1.5 mm} a}$ and $h^{a}$ is given by
\begin{equation}\label{39}
  \begin{split}
     L^{(2)}=-M_{p} \{ & \sigma (u_{1} \cdot \bar{D} v_{1}+ \frac{1}{2} \bar{e} \cdot v_{1} \times v_{1} + \frac{1}{2 l^{2}} \bar{e} \cdot u_{1} \times u_{1} ) \\
       & + \gamma (u_{2} \cdot \bar{D} v_{2}+ \frac{1}{2} \bar{e} \cdot v_{2} \times v_{2} + \frac{1}{2 l^{2}} \bar{e} \cdot u_{2} \times u_{2} ) \\
       & + \frac{m^{2}}{2} \beta _{1} \gamma \bar{e} \cdot (u_{1}-u_{2}) \times (u_{1}-u_{2}) -\frac{1}{2 \mu} v_{1} \cdot \bar{D} v_{1} \\
       & - \frac{1}{2 \mu l^{2}} u_{1} \cdot \bar{D} u_{1} - \frac{1}{\mu l^{2}} \bar{e} \cdot u_{1} \times v_{1} - u_{1} \cdot \bar{D} p - \bar{e} \cdot v_{1} \times p \},
  \end{split}
\end{equation}
where $\bar{D}$ is exterior covariant derivative with respect to $\bar{\omega}$. Now using \eqref{38} we linearize the equations of motion \eqref{4}-\eqref{8}, or equivalently, one can extract the linearized equations of motion from the Lagrangian \eqref{39}, and we will have
\begin{equation}\label{40}
\begin{split}
     & \bar{D} u_{1} + \bar{e} \times v_{1}=0, \\
     & \bar{D} u_{2} + \bar{e} \times v_{2}=0, \\
     & \bar{D} v_{1} + \frac{1}{l^{2}} \bar{e} \times u_{1} + \mu \bar{e} \times p =0, \\
     & \bar{D} v_{2} + \frac{1}{l^{2}} \bar{e} \times u_{2} - m^{2} \beta _{1} \bar{e} \times (u_{1}-u_{2}) =0, \\
     & \bar{D} p + \mu \sigma \bar{e} \times p - m^{2} \beta _{1} \gamma \bar{e} \times (u_{1}-u_{2}) =0,
\end{split}
\end{equation}
These linearized equations reduce to the result for topologically massive gravity theory when we take $u_{2}=v_{2}=\beta _{1}=\gamma =0$ (one can check this by looking the equation (3.8) in \cite{3} with $\alpha=0$). \\
Now, we introduce following transformations from $(u_{1},u_{2},v_{1},v_{2},p)$ to new Lorentz vector valued one-form fluctuations $(u_{+},u_{-},q_{1},q_{2},q_{3})$:
\begin{equation}\label{41}
\begin{split}
     & u_{+}=x_{+} \left( \frac{(\sigma \mu l - 1)}{\mu l^{2}} u_{1} +\frac{\gamma}{l} u_{2} +\frac{(\sigma \mu l - 1)}{\mu l} v_{1}+ \gamma v_{2} - p \right) \\
     & u_{-}=x_{-} \left( \frac{(\sigma \mu l + 1)}{\mu l^{2}} u_{1} +\frac{\gamma}{l} u_{2} -\frac{(\sigma \mu l + 1)}{\mu l} v_{1}- \gamma v_{2} + p \right) \\
     &  q_{1}=x_{1} \left( \lambda _{1} u_{1} - \lambda _{1} u_{2} + v_{1} - v_{2} - \frac{\mu }{(\mu \sigma - \lambda _{1})} p \right) \\
     & q_{2}=x_{2} \left( \lambda _{2} u_{1} - \lambda _{2} u_{2} + v_{1} - v_{2} - \frac{\mu }{(\mu \sigma - \lambda _{2})} p \right) \\
     & q_{3}=x_{3} \left( \lambda _{3} u_{1} - \lambda _{3} u_{2} + v_{1} - v_{2} - \frac{\mu }{(\mu \sigma - \lambda _{3})} p \right)
\end{split}
\end{equation}
where $\lambda _{1}$, $\lambda _{2}$ and $\lambda _{3}$ are the roots of following cubic equation,
\begin{equation}\label{42}
\lambda ^{3} - \mu \sigma \lambda ^{2} -(\frac{1}{l^{2}} + m^{2} \beta _{1}) \lambda + (\frac{\mu \sigma}{l^{2}} +m^{2} \beta _{1} \mu \sigma +\gamma \mu m^{2} \beta _{1} ) = 0,
\end{equation}
also $(x_{+},x_{-},x_{1},x_{2},x_{3})$ are arbitrary constants. We demand that all of these roots are real, so we must impose the following condition ( e.g. use the Tschirnhous-Vieta method of solving the cubic equation)
\begin{equation}\label{43}
\frac{1}{l^{2}} + m^{2} \beta _{1} + \frac{1}{3} \mu ^{2} \sigma ^{2} \geq  0.
\end{equation}
By using the transformations \eqref{41} we can diagonalize the linearized equations \eqref{40} as follow
\begin{equation}\label{44}
\begin{split}
     & \bar{D} u_{+} + \frac{1}{l} \bar{e} \times u_{+}=0, \\
     & \bar{D} u_{-} - \frac{1}{l} \bar{e} \times u_{-}=0, \\
     & \bar{D} q_{1} + \lambda _{1} \bar{e} \times q_{1}=0, \\
     & \bar{D} q_{2} + \lambda _{2} \bar{e} \times q_{2}=0, \\
     & \bar{D} q_{3} + \lambda _{3} \bar{e} \times q_{3}=0.
\end{split}
\end{equation}
In this manner, we can expect that the transformations \eqref{41} diagonalize the Lagrangian \eqref{39}. So we can rewrite the Lagrangian \eqref{39} in the diagonalized form, in terms of new 1-form fields as follow
\begin{equation}\label{45}
\begin{split}
   -\frac{L^{(2)}}{M_{p}}= \{ & A_{+} (u_{+} \cdot \bar{D} u_{+} + \frac{1}{l} \bar{e} \cdot u_{+} \times u_{+})+A_{-} (u_{-} \cdot \bar{D} u_{-} - \frac{1}{l} \bar{e} \cdot u_{-} \times u_{-}) \\
     & + B_{1} \lambda_{1} (q_{1} \cdot \bar{D} q_{1} + \lambda _{1} \bar{e} \cdot q_{1} \times q_{1}) + B_{2} \lambda_{2} (q_{2} \cdot \bar{D} q_{2} + \lambda _{2} \bar{e} \cdot q_{2} \times q_{2}) \\
     & + B_{3} \lambda_{3} (q_{3} \cdot \bar{D} q_{3} + \lambda _{3} \bar{e} \cdot q_{3} \times q_{3}) \},
\end{split}
\end{equation}
where
\begin{equation}\label{46}
\begin{split}
   & A_{+}=\frac{\mu l^{2}}{4 x_{+}^{2} (\sigma \mu l + \gamma \mu l -1)}, \hspace{1 cm} A_{-}=-\frac{\mu l^{2}}{4 x_{-}^{2} (\sigma \mu l + \gamma \mu l +1)}, \\
     & B_{1}=-\frac{\gamma (\mu \sigma - \lambda _{1}) [ (\gamma +\sigma) \mu l^{2} (m^{2} \beta _{1} + \mu \sigma \lambda _{1} - \lambda _{1} ^{2} )+ \mu \sigma - \lambda _{1}]}{2 x_{1}^{2} \lambda _{1}^{2} (\sigma \mu l + \gamma \mu l -1) (\sigma \mu l + \gamma \mu l +1) (\lambda _{3} - \lambda _{1})(\lambda _{2} - \lambda _{1}) }, \\
     & B_{2}=-\frac{\gamma (\mu \sigma - \lambda _{2}) [ (\gamma +\sigma) \mu l^{2} (m^{2} \beta _{1} + \mu \sigma \lambda _{2} - \lambda _{2} ^{2} )+ \mu \sigma - \lambda _{2}]}{2 x_{2}^{2} \lambda _{2}^{2} (\sigma \mu l + \gamma \mu l -1) (\sigma \mu l + \gamma \mu l +1) (\lambda _{3} - \lambda _{2})(\lambda _{1} - \lambda _{2}) }, \\
     & B_{3}=-\frac{\gamma (\mu \sigma - \lambda _{3}) [ (\gamma +\sigma) \mu l^{2} (m^{2} \beta _{1} + \mu \sigma \lambda _{3} - \lambda _{3} ^{2} )+ \mu \sigma - \lambda _{3}]}{2 x_{3}^{2} \lambda _{3}^{2} (\sigma \mu l + \gamma \mu l -1) (\sigma \mu l + \gamma \mu l +1) (\lambda _{1} - \lambda _{3})(\lambda _{2} - \lambda _{3}) }.
\end{split}
\end{equation}
Two first terms in the above Lagrangian can be written in the form of the difference of two linearized $SL(2,R)$ Chern-Simons 3-forms, so the $u_{\pm}$ fields have no local degrees of freedom. According to the analysis carried out in \cite{3} the Lagrangian which describe a single spin-2 mode of helicity $\pm 2$ has following form
\begin{equation}\label{47}
  L_{q}=-A M (q \cdot \bar{D} q + M \bar{e} \cdot q \times q),
\end{equation}
so that the Fierz-Pauli mass is given by $\mathcal{M}^{2}=M^{2}-1/l^{2}$, and the no-ghost condition for this Lagrangian 3-form can imposed as $A > 0$.
Then, each of the three last terms in \eqref{45} describe a single spin-2 mode of helicity $\pm 2$ and they are not ghost as long as $B_{i} > 0$(i=1,2,3). We guess that two of these conditions will be trivial but the third, say $B_{3} >0 $, impose a condition on parameters of the theory for which the Lagrangian \eqref{45} describe a theory with 6 physical degrees of freedom.

\section{AdS wave solution}
In this section, we are looking for AdS waves propagating on AdS$_{3}$ background. As we know, the AdS$_{3}$ metric in Poincare coordinates is
\begin{equation}\label{48}
  d \bar{s}^{2}= \frac{l^{2}}{y^{2}} (-2 du dv +dy^{2}).
\end{equation}
Now we consider the Kerr-Schild deformation of AdS$_{3}$ space-time as
\begin{equation}\label{49}
  g_{\mu \nu}= \bar{g}_{\mu \nu} - f(u,y) k_{\mu} k_{\nu} ,
\end{equation}
where the $f(u,y)$ is the wave profile and $k^{\mu}$ is a null, tangent to geodesic vector field. If we suppose that $k= \frac{y}{l} \partial _{v}$ then we have
\begin{equation}\label{50}
  ds^{2}= \frac{l^{2}}{y^{2}} (- f(u,y) du^{2}-2 du dv +dy^{2}) ,
\end{equation}
We will choose the following dreibein for this metric:
\begin{equation}\label{51}
  e_{1}^{\hspace{1.5 mm \hat{0}}}=\frac{l}{y} (\sqrt{f} du +\frac{dv}{\sqrt{f}}) , \hspace{1 cm} e_{1}^{\hspace{1.5 mm \hat{1}}}=\frac{l}{y} \frac{dv}{\sqrt{f}},\hspace{1 cm} e_{1}^{\hspace{1.5 mm \hat{2}}}=\frac{l}{y} dy,
\end{equation}
where the hat on numbers refers to Lorentz indices. Using \eqref{9} and \eqref{37}, we calculate $e_{2}$ and we have
\begin{equation}\label{52}
  e_{2}^{\hspace{1.5 mm \hat{0}}}= g(u,y) du + h(u,y) dv , \hspace{5 mm} e_{2}^{\hspace{1.5 mm \hat{1}}}=p(u,y) du + q(u,y) dv,\hspace{5 mm} e_{2}^{\hspace{1.5 mm \hat{2}}}=s(u,y) dy,
\end{equation}
where
\begin{equation}\label{53}
  \begin{split}
       & g(u,y)= \frac{2 \beta _{1} \gamma m^{2} l^{2} f(u,y) + \sigma y  ( \frac{\partial}{\partial y} - y \frac{\partial ^{2}}{\partial y ^{2}} ) f(u,y) - \frac{1}{\mu l} y^{3} \frac{\partial ^{3} f(u,y)}{\partial y ^{3}}}{2 \beta _{1} m^{2} l y \sqrt{f(u,y)} }, \\
       & h(u,y)=q(u,y)= \frac{\gamma l}{y \sqrt{f(u,y)}}, \hspace{1 cm} s(u,y)=\frac{\gamma l}{y}, \\
       & p(u,y)= \frac{ \sigma ( \frac{\partial}{\partial y} - y \frac{\partial ^{2}}{\partial y ^{2}} ) f(u,y) - \frac{1}{\mu l} y^{2} \frac{\partial ^{3} f(u,y)}{\partial y ^{3}}}{2 \beta _{1} m^{2} l \sqrt{f(u,y)} } .
  \end{split}
\end{equation}
Using \eqref{51} we can calculate \eqref{12} and \eqref{13} and we obtain
\begin{equation}\label{54}
  \Omega^{(2n)a}_{\hspace{7 mm} \mu}= -\frac{y^{3}}{\mu \sigma l^{2}} \left( \frac{\sigma}{\alpha _{1} l^{2}}\right) ^{n} \left( (2 + \sigma \mu l ) \frac{\partial ^{3} f(u,y)}{\partial y ^{3}} + y \frac{\partial ^{4} f(u,y)}{\partial y ^{4}} \right)k_{\mu} k^{a},
\end{equation}
where $ n \geq 1 $. By substituting the above expression into \eqref{11} we will have
\begin{equation}\label{55}
\omega ^{ \hspace{1 mm} a} _{2 \hspace{1.5 mm} \mu} = \omega ^{ \hspace{1 mm} a} _{1 \hspace{1.5 mm} \mu} + \frac{y^{3}}{\mu l^{2} (\sigma - \alpha _{1} m^{2} l^{2} )}  \left( (2 + \sigma \mu l ) \frac{\partial ^{3} f(u,y)}{\partial y ^{3}} + y \frac{\partial ^{4} f(u,y)}{\partial y ^{4}} \right)k_{\mu} k^{a} ,
\end{equation}
Now using above results, we can calculate $R_{2}$ and then by substituting these results into \eqref{6} we will have the following fifth order differential equation for the wave profile:
\begin{equation}\label{56}
\begin{split}
     & \sigma \mu l (1+ M_{0}^{2} l^{2}) \left( \frac{\partial f(u,y)}{\partial y } - y \frac{\partial ^{2} f(u,y)}{\partial y ^{2}} \right)  \\
     & + [2 (\sigma \mu l +2) - \alpha _{2} \gamma ^{2} m^{2} l ^{2} ] y ^{2} \frac{\partial ^{3} f(u,y)}{\partial y ^{3}} \\
     & + (\sigma \mu l +5) y ^{3} \frac{\partial ^{4} f(u,y)}{\partial y ^{4}} + y ^{4} \frac{\partial ^{5} f(u,y)}{\partial y ^{5}}=0,
\end{split}
\end{equation}
where $M_{0}^{2}= (\gamma+ \sigma) m^{2} \beta _{1} / \sigma$. We know that such equations has the polynomial solutions, so we take $f(u,y)=\tilde{f}(u) y^{n}$ and substitute this expression into \eqref{56} then we obtain following quintic equation for $n$,
\begin{equation}\label{57}
\begin{split}
     &  n (n-2) \{ -\sigma \mu l (1+ M_{0}^{2} l^{2}) + [2 (\sigma \mu l +2) - \alpha _{2} \gamma ^{2} m^{2} l ^{2} ] (n-1) \\
     & + (\sigma \mu l +5) (n-1)(n-3) + (n-1)(n-3) (n-4) \} = 0.
\end{split}
\end{equation}
One can solve this equation for obtaining $n$, and hence the generic solution for the wave profile is:
\begin{equation}\label{58}
  f(u,y)=f_{0}(u)+f_{2}(u) y^{2}+f_{N_{1}}(u) y^{1+N_{1}}+f_{N_{2}}(u) y^{1+N_{2}}+f_{N_{3}}(u) y^{1+N_{3}},
\end{equation}
where $N_{1}$, $N_{2}$ and $N_{3}$ are the roots of the following cubic equation,
\begin{equation}\label{59}
 -\sigma \mu l (1+ M_{0}^{2} l^{2}) - \alpha _{2} \gamma ^{2} m^{2} l ^{2} N + \sigma \mu l N^{2}+N^{3}= 0.
\end{equation}
One can easily check that all the results of this section will reduce to ZDG one (with $\beta_{2}=0$) \cite{4} when we tends $\mu$ to infinity.

\section{Conclusion}
In this paper, we have considered a generalization of the ZDG with $\beta _{2}=0$. We have added a Chern-Simons term to the Lagrangian of ZDG with $\beta _{2}=0$ and, in addition, for fixing the torsion associated to $e_{1}$ and $\omega _{1}$ we have added an extra term which is proportional to $h \cdot T_{1}$. Using the field equations, we could obtain a differential equation for $e_{1}$ and this guarantee that this model is well-define. In the section 3, we alleged that this model is a Chern-Simons like theory of gravity and we have considered Hamiltonian analysis, which provided in \cite{2}, for counting local degrees of freedom. It is important that by a Hamiltonian analysis one can obtain the number of local degrees of freedom exactly and independent of a linearised approximation. We showed that this model have 6  phase space degrees of freedom and we deduced that one can avoided from ghost with imposing some conditions on the parameters of the theory. We have obtained the quadratic Lagrangian for the perturbations about AdS$_3$ background in section 4. From quadratic Lagrangian $L^{(2)}$ in Eq.(\ref{45}), one can see that the no ghost conditions for 3 spin-2 modes are $B_{i}>0, (i=1,2,3)$. As have been shown in \cite{2} the total dimension of physical phase space of GZDG is 4 and it propagate two spin-2 modes with different masses. It is interesting that GZDG$^+$ propagate 3 spin-2 modes with different masses. Very recently it has been shown that GZDG can be reduce to GMMG \cite{set} \footnote{Generalized Minimal Massive Gravity (GMMG) which is a generalized version of Minimal Massive Gravity (MMG) \cite{3} is realized
by adding the CS deformation term, the higher derivative deformation term, and an extra term
to pure Einstein gravity with a negative cosmological constant.}. So may be GZDG$^+$ can reduce to another version of MMG.
Finally we have obtained AdS waves, which propagate on AdS$_{3}$ background, as an example solution for this model.

\section{Acknowledgments}
M. R. Setare thank  W. Merbis, A. J. Routh, and A. F. Goya, for helpful discussions and correspondence.


\begin{thebibliography}{9}
\bibitem{2'}S. Deser, R. Jackiw, and G. 't Hooft, "Three-dimensional einstein gravity: Dynamics
of flat space," Ann. Phys. 152, 220 (1984).
\bibitem{3'}S. Deser and R. Jackiw, "Three-dimensional cosmological gravity: Dynamics of
constant curvature," Annals Phys. 153, 405 (1984).
\bibitem{4'}S. Deser, R. Jackiw and S. Templeton, Annals Phys.
140, 372 (1982) [Erratum-ibid. 185, 406.1988 APNYA,
281,409 (1988 APNYA,281,409-449.2000)].
\bibitem{6}E. A. Bergshoeff, O. Hohm and P. K. Townsend, Phys. Rev. Lett. 102, 201301, (2009).
\bibitem{d1}M. Nakasone and I. Oda, Prog. Theor. Phys. 121, 1389, (2009).
\bibitem{d2} M. Nakasone and I. Oda, Phys. Rev. D79, 104012, (2009).
\bibitem{dan11}D. Grumiller and N. Johansson, JHEP 0807, 134 (2008)
\bibitem{dan12} S. Ertl, D. Grumiller and N. Johansson, arXiv:0910.1706 [hep-th].
\bibitem{oh1}K. Muneyuki and N. Ohta, Phys. Rev. D 85, 101501, (2012).
\bibitem{9} A. Sinha, JHEP 1006, 061, (2010) .
\bibitem{10}R. C. Myers and A. Sinha, JHEP 1101, 125, (2011).
\bibitem{11}S. Deser and B. Tekin, Class. Quant. Grav. 20, L259, (2003).
\bibitem{12}I. Gullu, T. C. Sisman and B. Tekin, Class. Quant. Grav. 27, 162001, (2010).
\bibitem{13}M. F. Paulos, Phys. Rev. D 82, 084042, (2010).
\bibitem{1}E. A. Bergshoeff, S. de Haan, O. Hohm, W. Merbis, and P. K. Townsend,
Phys. Rev. Lett. 111, 111102, (2013).
 \bibitem{2}E. A. Bergshoeff, O. Hohm, W. Merbis, A. J. Routh and P. K. Townsend,
Lect. Notes Phys. 892, 181, (2015).
\bibitem{4} Eric A. Bergshoeff, Andres F. Goya, Wout Merbis, Jan Rosseel, JHEP 04 (2014) 012.
\bibitem{27}M. Ba˜nados, C. Deffayet,  M. Pino, Phys. Rev. D 88, 124016, (2013).
\bibitem{set}M. R. Setare, Nucl. Phys. B898, 259, (2015).
\bibitem{3} Eric Bergshoeff, Olaf Hohm, Wout Merbis, Alasdair J. Routh, Paul K. Townsend, Class. Quant. Grav. 31, 145008, (2014).

\end{thebibliography}
\end{document}